\documentclass[numbers]{elsarticle}
\usepackage{latexsym}
\usepackage{natbib,url}
\usepackage[latin2]{inputenc}
\makeatletter
\def\ps@pprintTitle{%
	\let\@oddhead\@empty
	\let\@evenhead\@empty
	\def\@oddfoot{\centerline{\thepage}}%
	\let\@evenfoot\@oddfoot}
\makeatother

\begin{document}

\title{Parameterizable Consensus Connectomes from the Human Connectome Project: The Budapest Reference Connectome Server v3.0}
	
\author[p]{Balázs Szalkai\corref{cor2}}
\ead{szalkai@pitgroup.org}
\author[p]{Csaba Kerepesi\corref{cor2}}
\ead{kerepesi@pitgroup.org}
\author[p]{Bálint Varga\corref{cor2}}
\ead{balorkany@pitgroup.org}
\author[p,u]{Vince Grolmusz\corref{cor1}}
\ead{grolmusz@pitgroup.org}
\cortext[cor1]{Corresponding author}
\cortext[cor2]{Joint first authors}
\address[p]{PIT Bioinformatics Group, Eötvös University, H-1117 Budapest, Hungary}
\address[u]{Uratim Ltd., H-1118 Budapest, Hungary}

\date{}

\begin{abstract}

Connections of the living human brain, on a macroscopic scale, can be mapped by a diffusion MR imaging based workflow. Since the same anatomic regions can be corresponded between distinct brains, one can compare the presence or the absence of the edges, connecting the very same two anatomic regions, among multiple cortices. Previously, we have constructed the consensus braingraphs on 1015 vertices first in five, then in 96 subjects in the Budapest Reference Connectome Server v1.0 and v2.0, respectively. Here we report the construction of the version 3.0 of the server, generating the common edges of the connectomes of variously parameterizable subsets of the 1015-vertex connectomes of 477 subjects of the Human Connectome Project's 500-subject release. The consensus connectomes are downloadable in csv and GraphML formats, and they are also visualized on the server's page. The consensus connectomes of the server can be considered as the ``average, healthy'' human connectome since all of their connections are present in at least $k$ subjects, where the default value of $k=209$, but it can also be modified freely at the web server.
The webserver is available at \url{http://connectome.pitgroup.org}.
\end{abstract}

\maketitle

\section{Introduction} 

A dermatologist recognizes the healthy skin by comparing it to examples from the medical praxis and the literature. Similarly, a pathologist may distinguish healthy and diseased tissues by using on-line case databases (e.g., at \url{http://path.upmc.edu/cases.html}). Neuroscientists, until now, have no similar resource for describing the neural connections in the human brain that are present in most of the healthy subjects. These ``frequently appearing'' connections, the consensus connectomes, can also be applied for error-correction: by setting the parameters properly in the webserver, one can get rid of the connections that were erroneously detected in a few subjects, either due to some measurement or computational artifact.

 Here we describe the Budapest Reference Connectome Server v3.0, a webserver, which is capable of generating the consensus connectomes from the diffusion MRI data of healthy subjects, collected and deposited in the Human Connectome Project \cite{McNab2013}. The consensus connectomes can be prepared by the users, setting several parameters and can also be visualized at the site or with any other software, handling GraphML formats.
 
 We represent cerebral connections as graphs: the vertices correspond to 1015 regions of interests (ROIs) of the gray matter, and two vertices (corresponding to two of these ROIs) are connected by an edge if tractography finds at least one neuronal fiber tract between them. 

Version 1.0 of the Budapest Reference Connectome Server was prepared  from six connectomes of five subjects, based on the data published in \cite{Hagmann2008}. Version 2.0 of the webserver \cite{Szalkai2015a} was compiled from 96 connectomes, computed from the Human Connectome Project's \cite{McNab2013} 500-subjects release. We have reported a surprising and unforeseen discovery, found by changing the parameters of the version 2.0 of the webserver in \cite{Kerepesi2015}.

\begin{figure}[h!]
	\centering
	\includegraphics[width=80mm]{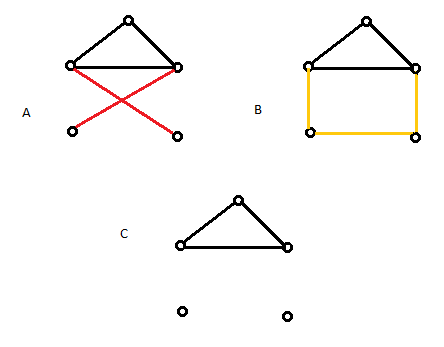}
	\caption{An example for the common edges in graphs A and B. In the graphs A and B the black edges are common; the red edges are present only in graph A, the yellow ones only in graph B. Graph C contains the common (or consensus) edges of graphs A and B.}
\end{figure}

The main feature of the server is the identification of the graph edges, which are present in at least $k$ connectomes, between the same gray matter areas (see Figure 1 for an example of common edges of graphs). Here we describe the application and the capabilities of the webserver, and we also describe its construction. The webserver is available at \url{http://connectome.pitgroup.org}.

\section{Results and Discussion}

Here we describe version 3.0 of the webserver; versions 1.0 and 2.0 (that are also available at \url{http://connectome.pitgroup.org}) were described in  \cite{Szalkai2015a}. 

The default consensus graph, called ``Budapest Reference Connectome v3.0'' can be downloaded  directly, without changing any settings, by clicking on the ``Download graph'' link on the upper right corner of the graphical user interface. Here we detail the selectable parameters, and also give their default settings, which lead to the default reference connectome.

\begin{figure}[h!]
	\centering
	\includegraphics[width=120mm]{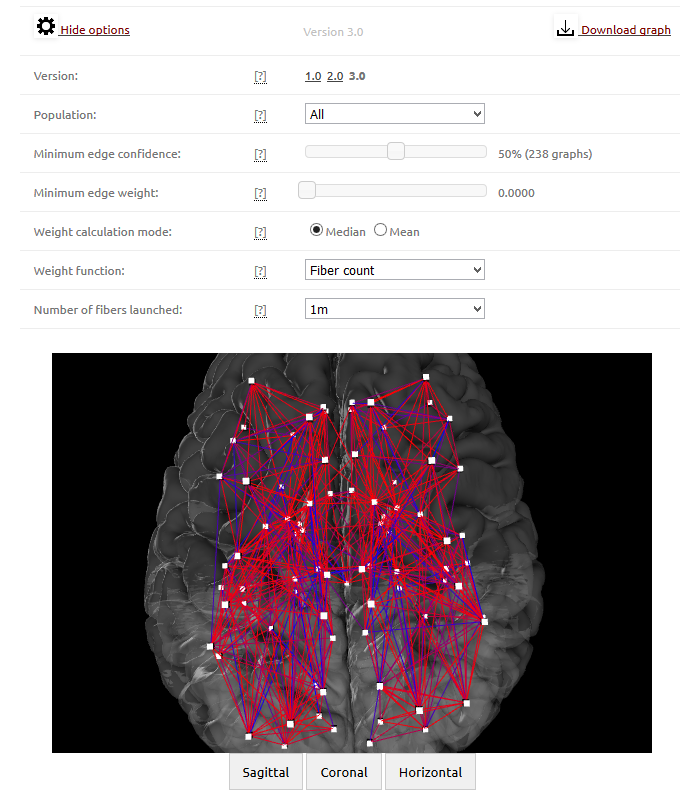}
	\caption{The screenshot of the Budapest Reference Connectome Server v3.0.}
\end{figure}

After the "Show options" button is clicked, several options can be set. Right to each option, a question mark [?] can be seen that succinctly describes the option in question. 
\begin{itemize}
	
	\item[(i)] Versions can be selected as 1.0, 2.0 or 3.0. The default choice is 3.0. 
	
	\item[(ii)] Population: The subjects can be filtered according to their sex: All, Male or Female. The default choice is ``All''.
	
	\item[(iii)] Minimum edge confidence: Only those edges are selected in the graph that are present in at least $p\%$ of the connectomes. Value $p$ can be chosen from 1 through 100. The default choice is 50\%. The webserver also gives the number of the connectomes, corresponding to the percentage selected. Note that depending on the last selectable parameter - ``Number of fibers launched `` -- the number of the connectomes may differ for the same percentage. 
	
	\item[(iv)] Minimum edge weight: Only those edges are included in the graph, whose mean or median weight is at least the value specified. The weight function and the ``mean'' or ``median'' computational methods are selectable. The default value is 0.
	
	\item[(v)] Weight calculation mode: Each edge in each connectome has a weight. If the ``Median'' option is chosen then the median of the different weights is computed for each edge, if the ``Mean'' option is selected then the mean, i.e., the average of the weights is computed for each edge. If this value is larger than or equal to the threshold in (iv), then the edge will be present in the consensus graph, otherwise, it will be left out of the graph. The default choice is ``Median'', since this selection is more robust in the case of few out-lier values.
	
	\item[(vi)] Weight function: Four weight functions can be chosen:
	
	\begin{itemize}
	
	\item[(a)] Electrical connectivity: for each edge, the number of fibers is divided by the average fiber length, defining the edge.
	
	\item[(b)] Fiber count: The number of fibers found between the two nodes, corresponding two ROIs.
	
	\item[(c)] Fiber length: The average length of fibers, connecting the two ROIs that correspond to the vertices of the edge.
	
	\item[(d)] Fractional anisotropy: The average of the fractional anisotropies of the fiber tracts \cite{Basser2011}.
	
	\end{itemize}
	
	The default weight function is ``Fiber count''.
	
	\item[(vii)] Number of fibers launched: In the tractography phase of the workflow, the number of the fibers that were launched. There are three choices: 20 thousand, 200 thousand and one million. The default is 20 thousand. 
	
	\end{itemize}
	
The graphical component on the website visualizes the graph, corresponding to the options selected. The graph can be rotated and magnified, and if the cursor is placed above a node, then its anatomical name appears. For a cleaner view, the graphical component sticks together the vertices that are situated close to one another; the downloadable graphs in CSV or GraphML formats, however, contain all the 1015 nodes in each graph generated.

The consensus graphs that were generated by the server can be downloaded in CSV or GraphML formats for further processing or visualization in other imaging software.

\section{Methods}

Our data source was the Human Connectome Project 500 Subjects Release (http://www.humanconnectome.org/documentation/S500/). We have applied the Connectome Mapper Toolkit \url{http://cmtk.org} for partitioning, tractography and graph construction. For tractography, the deterministic streamline method was chosen with 20 thousand, 200 thousand and one million fibers initiated.

The graphs computed are deposited at the website \url{http://braingraph.org/download-pit-group-connectomes/}.
 
After the graphs had been constructed, their edges with their edge confidences and edge weights were calculated and stored in pre-computed tables. These pre-computed tables were integrated into the webserver.

A modified version of the WebGL Brain Viewer \cite{Ginsburg2011} was applied in the visualization component of our server.

\section{Acknowledgments}
Data were provided in part by the Human Connectome Project, WU-Minn Consortium (Principal Investigators: David Van Essen and Kamil Ugurbil; 1U54MH091657) funded by the 16 NIH Institutes and Centers that support the NIH Blueprint for Neuroscience Research; and by the McDonnell Center for Systems Neuroscience at Washington University.


\bibliography{v:/vince/cikkek/connectome}

\begin{thebibliography}{7}
\providecommand{\natexlab}[1]{#1}
\providecommand{\url}[1]{\texttt{#1}}
\expandafter\ifx\csname urlstyle\endcsname\relax
  \providecommand{\doi}[1]{doi: #1}\else
  \providecommand{\doi}{doi: \begingroup \urlstyle{rm}\Url}\fi

\bibitem[McNab et~al.(2013)McNab, Edlow, Witzel, Huang, Bhat, Heberlein,
  Feiweier, Liu, Keil, Cohen-Adad, Tisdall, Folkerth, Kinney, and
  Wald]{McNab2013}
Jennifer~A. McNab, Brian~L. Edlow, Thomas Witzel, Susie~Y. Huang, Himanshu
  Bhat, Keith Heberlein, Thorsten Feiweier, Kecheng Liu, Boris Keil, Julien
  Cohen-Adad, M~Dylan Tisdall, Rebecca~D. Folkerth, Hannah~C. Kinney, and
  Lawrence~L. Wald.
\newblock The {H}uman {C}onnectome {P}roject and beyond: initial applications
  of 300 m{T}/m gradients.
\newblock \emph{Neuroimage}, 80:\penalty0 234--245, Oct 2013.
\newblock \doi{10.1016/j.neuroimage.2013.05.074}.
\newblock URL \url{http://dx.doi.org/10.1016/j.neuroimage.2013.05.074}.

\bibitem[Fischl(2012)]{Fischl2012}
Bruce Fischl.
\newblock Freesurfer.
\newblock \emph{Neuroimage}, 62\penalty0 (2):\penalty0 774--781, 2012.

\bibitem[Hagmann et~al.(2008)Hagmann, Cammoun, Gigandet, Meuli, Honey, Wedeen,
  and Sporns]{Hagmann2008}
Patric Hagmann, Leila Cammoun, Xavier Gigandet, Reto Meuli, Christopher~J.
  Honey, Van~J. Wedeen, and Olaf Sporns.
\newblock Mapping the structural core of human cerebral cortex.
\newblock \emph{PLoS Biol}, 6\penalty0 (7):\penalty0 e159, Jul 2008.
\newblock \doi{10.1371/journal.pbio.0060159}.
\newblock URL \url{http://dx.doi.org/10.1371/journal.pbio.0060159}.

\bibitem[Szalkai et~al.(2015)Szalkai, Kerepesi, Varga, and
  Grolmusz]{Szalkai2015a}
Bal{\'a}zs Szalkai, Csaba Kerepesi, B{\'a}lint Varga, and Vince Grolmusz.
\newblock The {B}udapest {R}eference {C}onnectome {S}erver v2. 0.
\newblock \emph{Neuroscience Letters}, 595:\penalty0 60--62, 2015.

\bibitem[Kerepesi et~al.(2015)Kerepesi, Szalkai, Varga, and
  Grolmusz]{Kerepesi2015}
Csaba Kerepesi, Bal{\'a}zs Szalkai, B{\'a}lint Varga, and Vince Grolmusz.
\newblock Does the budapest reference connectome server shed light to the
  development of the connections of the human brain?
\newblock \emph{arXiv preprint arXiv:1509.05703}, 2015.

\bibitem[Basser and Pierpaoli(1996)]{Basser2011}
Peter~J. Basser and Carlo Pierpaoli.
\newblock Microstructural and physiological features of tissues elucidated by
  quantitative-diffusion-tensor mri.
\newblock \emph{J Magn Reson}, 213\penalty0 (2):\penalty0 560--570, Dec 1996.
\newblock \doi{10.1016/j.jmr.2011.09.022}.
\newblock URL \url{http://dx.doi.org/10.1016/j.jmr.2011.09.022}.

\bibitem[Ginsburg et~al.(2011)Ginsburg, Gerhard, Congote, and
  Pienaar]{Ginsburg2011}
Daniel Ginsburg, Stephan Gerhard, John~Edgar Congote, and Rudolph Pienaar.
\newblock Realtime visualization of the connectome in the browser using webgl.
\newblock \emph{Frontiers in Neuroinformatics}, 2011.
\newblock URL
  \url{http://www.frontiersin.org/10.3389/conf.fninf.2011.08.00095/event\_abstract}.

\end{thebibliography}
\bibliographystyle{unsrtnat}

\end{document}